\documentclass[12pt,american]{article}
\usepackage{tgtermes}
\usepackage{helvet}
\usepackage[lf]{FiraMono}
\usepackage[T1]{fontenc}
\usepackage[utf8]{inputenc}
\usepackage[a4paper]{geometry}
\usepackage{verbatim}
\usepackage{amsmath}
\usepackage{amsthm}
\usepackage{amssymb}
\usepackage[authoryear]{natbib}

\makeatletter
\numberwithin{equation}{section}
\numberwithin{figure}{section}

\usepackage{pdfpages}

\usepackage{amsthm}

\theoremstyle{definition}
\newtheorem{assumption}{Assumption}

\theoremstyle{plain}
\newtheorem{proposition}{Proposition}
\newtheorem{theorem}{Theorem}
\newtheorem{corollary}{Corollary}

\makeatother

\usepackage{babel}
\begin{document}
\title{A Theory of Bootstrap Coverage Calibration for Generalized Posterior
Credible Sets}
\author{Masahiro Tanaka\thanks{Faculty of Economics, Fukuoka University, Fukuoka, Japan. Address:
8-19-1, Nanakuma, Jonan, Fukuoka, Japan 814-0180. E-mail: m.tanaka.tt@fukuoka-u.ac.jp.}}
\maketitle
\begin{abstract}
Generalized posteriors replace the likelihood by an exponentiated
empirical criterion, but their credible sets generally lack asymptotic
justification for frequentist coverage. General posterior calibration
selects a scalar learning rate by estimating coverage with the bootstrap.
Using Edgeworth expansions under regular fixed-dimensional asymptotics,
we derive higher-order coverage expansions and analyze the stochastic
approximation step used in the implemented algorithm. For a fixed
nominal level, the root of the bootstrap coverage equation is consistent
under a uniform coverage approximation and local identification. The
higher-order expansions separate two sources of coverage error: the
sampling Edgeworth correction for the estimator and the posterior
Edgeworth correction for credible set boundaries, centres, and shapes.
A scalar learning rate can calibrate all nominal levels in the Gaussian
limit only when the posterior covariance and the sampling covariance
are proportional. Hence, bootstrap calibration is a level-specific
scale correction, not a remedy for general shape misspecification.
\end{abstract}

\section{Introduction}

A generalized posterior replaces the likelihood using an exponential
transformation of an empirical criterion \citep{Martin2022}. Such
distributions arise when the target is defined by a loss, estimating
equation, pseudolikelihood, composite likelihood, or extremum criterion,
rather than a full sampling model. They retain posterior-like advantages,
such as the incorporation of prior information, simulation-based uncertainty
summaries, and direct credible set construction. The ordinary Bayesian
posterior is recovered when this criterion is the negative average
log-likelihood, and the learning rate is fixed at one.

However, this flexibility comes at a cost. For a correctly specified
likelihood model, the curvature determines both the posterior covariance
and the sampling covariance. For a general criterion, these roles
are separate. The generalized posterior is centred to first order
at the empirical minimizer, and its covariance is determined by the
inverse curvature of the criterion multiplied by the inverse learning
rate. The sampling covariance of the minimizer is a sandwich covariance
involving both the curvature and the variability of the empirical
scores. Unless these covariance matrices are proportional, a scalar
learning rate changes the size but not the shape. Consequently, nominal
posterior probability does not necessarily imply frequentist coverage.
See \citet{Miller2021} for a general asymptotic treatment of the
concentration, normal approximation, and coverage of generalized posteriors.

This calibration problem has been reported in several studies. Quasi-Bayesian
and Laplace-type approaches use exponentiated objective functions
to perform simulation-based inference for extremum estimators \citep{Chernozhukov2003,Jiang2008}.
Gibbs posteriors and general Bayesian updating provide posterior updates
based on loss functions without requiring a full probability model
for the data \citep{Zhang2006a,Zhang2006b,Bissiri2016}. The misspecification
literature emphasizes the gap between the posterior curvature and
sandwich sampling variability \citep{Mueller2013,Li2024}. Proposed
responses include sandwich posterior adjustments \citep{Shaby2014}
and information-matching or risk-based learning rate rules \citep{Gruenwald2017,Holmes2017,Lyddon2019}.
These approaches clarify why calibration is required. However, they
usually target an asymptotic covariance or predictive/risk criterion
(see \citealp{Wu2023} for a comparison).

General posterior calibration (GPC), as proposed by \citet{Syring2019},
was motivated by the more direct goal of constructing credible sets
with approximately correct coverage under repeated sampling. For a
fixed nominal level $1-\alpha$, GPC chooses the learning rate by
repeatedly bootstrapping the data, recomputing the generalized posterior,
constructing the corresponding credible set, and solving the bootstrap
analogue of the coverage equation. Its appeal is that it directly
targets the reported credible region and requires only posterior simulation
and bootstrap resampling. Recent studies have introduced computational
improvements for GPC \citep{Tanaka2024,Tanaka2025}, as well as a
calibration method for variational posterior inference \citep{Onizuka2024}.

Despite this appeal, the theoretical status of GPC is not entirely
straightforward. The implemented algorithm contains several layers:
bootstrap approximations to an unknown population coverage curve,
posterior simulation within each bootstrap sample, Monte Carlo estimation
of the coverage curve, and a stochastic approximation or root-finding
step for the learning rate. Moreover, even if the bootstrap coverage
curve is known exactly, it is not clear which conditions guarantee
that its root is close to the population root. It is also not clear
how higher-order errors affect calibrated coverage. Sampling skewness
and kurtosis affect the distribution of the estimator, while posterior
skewness, posterior kurtosis and random credible set boundaries affect
the posterior probability statement. Finally, as the learning rate
is scalar, it is important to distinguish what can be repaired by
calibrating at one nominal level from what cannot be repaired across
an entire family of credible sets.

This study provides a regular fixed-dimensional asymptotic account
of these issues. The object studied is not the finite Monte Carlo
algorithm itself, but the exact bootstrap coverage equation towards
which it is directed. The results have three main implications. First,
exact bootstrap calibration is an inverse problem: under a uniform
coverage approximation and local identification, the bootstrap root
inherits this approximation rate. Second, a higher-order expansion
separates sampling Edgeworth error from the posterior boundary, centre,
and shape corrections. Third, the scalar learning rate can calibrate
all levels within the Gaussian limit only when the posterior curvature
covariance is proportional to the sampling covariance.

The results are restricted to regular fixed-dimensional problems and
credible sets admitting scalar boundaries or radial representations.
These include marginal credible intervals and asymptotically ellipsoidal
credible sets, but do not yield a full multivariate Edgeworth theory
for arbitrary highest posterior density sets. The finite bootstrap,
posterior Monte Carlo, and stochastic approximation errors of the
implemented GPC algorithm are separated from the statistical target
and are treated only at the local level.

\section{General Posterior Calibration}

Let $X_{1},\ldots,X_{n}$ be independent observations with law $P_{0}$.
Let $\theta\in\Theta\subset\mathbb{R}^{d}$. For a criterion $R_{n}(\theta)$
and a prior density $\pi(\theta)$, we define the generalized posterior
\[
\Pi_{n,\omega}\left(d\theta\mid X^{n}\right)=\frac{\exp\{-\omega nR_{n}(\theta)\}\pi(\theta)d\theta}{\int_{\Theta}\exp\{-\omega nR_{n}(t)\}\pi(t)dt},\qquad\omega>0.
\]
When $R_{n}$ is the negative average log-likelihood and $\omega=1$,
this is an ordinary posterior. For a general loss or risk, $\omega$
controls the posterior scale. Let $\theta_{0}=\theta(P_{0})$ be the
population target; for example, the minimizer of $R(\theta)=P_{0}\ell_{\theta}$,
and let $C_{n,\omega,\alpha}=C_{n,\omega,\alpha}(X^{n})$ be a posterior
credible set with posterior probability $1-\alpha$. Its frequentist
coverage is 
\[
c_{n,\alpha}(\omega)=P_{0}\{\theta_{0}\in C_{n,\omega,\alpha}\}.
\]
General posterior calibration seeks a solution to $c_{n,\alpha}(\omega)=1-\alpha$.
As $P_{0}$ and $\theta_{0}$ are unknown, they are replaced by a
bootstrap law $P_{n}^{*}$ and an estimator $\widehat{\theta}_{n}=\theta(\widehat{P}_{n})$.
If $X_{1}^{*},\ldots,X_{n}^{*}$ are drawn conditionally from $\widehat{P}_{n}$,
then let $R_{n}^{*}$ be the corresponding criterion, and let 
\[
\Pi_{n,\omega}^{*}(d\theta\mid X^{*n},X^{n})\propto\exp\{-\omega nR_{n}^{*}(\theta)\}\pi(\theta)d\theta.
\]
Let $C_{n,\omega,\alpha}^{*}(X^{*n})$ be the bootstrap credible set.
The bootstrap coverage curve for the exact bootstrap is 
\begin{equation}
c_{n,\alpha}^{*}(\omega)=P^{*}\{\widehat{\theta}_{n}\in C_{n,\omega,\alpha}^{*}(X^{*n})\},\label{eq:bootstrap-coverage}
\end{equation}
where $P^{*}$ denotes conditional probability given the original
data. The bootstrap-calibrated learning rate $\widehat{\omega}_{n}$
solves $c_{n,\alpha}^{*}(\widehat{\omega}_{n})=1-\alpha$. The stochastic
approximation used in practice approximates (\ref{eq:bootstrap-coverage})
by a Monte Carlo method, and this additional layer is considered in
Section 5.

\section{Regular Asymptotic Structure}

Let $\widehat{\theta}_{n}$ be the local minimizer around which the
generalized posterior concentrates. The regular theory uses the following
abstract regularity conditions.

\begin{assumption}[Local posterior and sampling limits]

The following statements hold.
\begin{enumerate}
\item Uniformly on $h=O(1)$, with $\theta=\widehat{\theta}_{n}+n^{-1/2}h$,
\[
n\{R_{n}(\theta)-R_{n}(\widehat{\theta}_{n})\}=\frac{1}{2}h^{\top}J_{n}h+n^{-1/2}A_{3n}(h)+n^{-1}A_{4n}(h)+o_{p}(n^{-1}),
\]
where the eigenvalues of $J_{n}$ are bounded away from zero and infinity
with probability tending to one. Conditionally on the original data,
the bootstrap criterion admits an analogous expansion around its bootstrap
local minimizer $\widehat{\theta}_{n}^{*}$: uniformly on $h=O(1)$,
with $\theta=\widehat{\theta}_{n}^{*}+n^{-1/2}h$,
\[
n\{R_{n}^{*}(\theta)-R_{n}^{*}(\widehat{\theta}_{n}^{*})\}=\frac{1}{2}h^{\top}J_{n}^{*}h+n^{-1/2}A_{3n}^{*}(h)+n^{-1}A_{4n}^{*}(h)+o_{P^{*}}(n^{-1}),
\]
on bootstrap-regular events whose conditional probability tends to
one. The eigenvalues of $J_{n}^{*}$ are bounded away from zero and
infinity with $P^{*}$-probability tending to one.
\item The prior is positive and sufficiently smooth near $\theta_{0}$,
and the posterior mass outside fixed local neighbourhoods is negligible.
Also, 
\[
n^{1/2}(\widehat{\theta}_{n}-\theta_{0})\rightsquigarrow N(0,\Psi),\qquad\Pi_{n,\omega}\{n^{1/2}(\theta-\widehat{\theta}_{n})\in\cdot\mid X^{n}\}\rightsquigarrow N(0,\omega^{-1}J^{-1}),
\]
where $J$ is the probability limit of $J_{n}$ and $\Psi$ is positive
definite.
\item For every compact interval $\Omega\subset(0,\infty)$ and every fixed
scalar contrast $a^{\top}\theta$, the original and bootstrap posteriors
admit posterior Edgeworth expansions of the type justified for a regular
Bayesian posterior by \citet{Kolassa2020}, uniformly for $\omega\in\Omega$
on regular events.
\end{enumerate}
\end{assumption}

One set of sufficient regularity conditions, together with the required
uniform Edgeworth assumptions, is provided in Sections S1 and S2 of
the Supplementary Material.

For multivariate credible ellipsoids or their first-order equivalents,
where $J_{n}$ is replaced by $J$, we consider 
\[
E_{n,\omega,\alpha}=\{\theta:n(\theta-\widehat{\theta}_{n})^{\top}J(\theta-\widehat{\theta}_{n})\le\frac{q_{d,1-\alpha}}{\omega}+o_{p}(1)\},
\]
where $q_{d,1-\alpha}$ is the $(1-\alpha)$-th quantile of $\chi_{d}^{2}$.

\begin{proposition}[Limits of scalar calibration]

Under Assumption 1, for each fixed nominal level, the first-order
coverage of $E_{n,\omega,\alpha}$ is:
\[
P\left\{ Q_{\Lambda}\le\frac{q_{d,1-\alpha}}{\omega}\right\} +o(1),\qquad Q_{\Lambda}=\sum_{j=1}^{d}\lambda_{j}Z_{j}^{2},
\]
where $Z_{j}$ are independent standard normal variables and $\lambda_{1},\ldots,\lambda_{d}$
are the eigenvalues of $J^{1/2}\Psi J^{1/2}$. A scalar $\omega_{0}$
calibrates every $\alpha\in(0,1)$ to first order if and only if
\begin{equation}
\lambda_{1}=\cdots=\lambda_{d}=\omega_{0}^{-1},\qquad\text{equivalently}\qquad\Psi=\omega_{0}^{-1}J^{-1}.\label{eq:proportionality}
\end{equation}

\end{proposition}

The proof is provided in Section S3 of the Supplementary Material.

Thus, scalar calibration is level-specific unless (\ref{eq:proportionality})
holds.

\begin{assumption}[Uniform bootstrap approximation and identification of the coverage root]

Fix $\alpha\in(0,1)$ and let $\Omega\subset(0,\infty)$ be compact.
For all sufficiently large $n$, suppose $c_{n,\alpha}$ has a unique
root $\omega_{n}^{\dagger}\in\textrm{int}\left(\Omega\right)$. There
exist a sequence $a_{n}\to0$ and convex neighbourhoods $N_{n}\subset\Omega$
of $\omega_{n}^{\dagger}$ such that the following conditions hold:
\begin{enumerate}
\item The population coverage curve $c_{n,\alpha}$ is continuous on $\Omega$
and satisfies $c_{n,\alpha}(\omega_{n}^{\dagger})=1-\alpha$.
\item The root is locally identified. For every sufficiently small $\varepsilon>0$,
there exists $\delta_{\varepsilon}>0$ such that, for all large $n$,
\[
\inf\{\left|c_{n,\alpha}(\omega)-(1-\alpha)\right|:\omega\in\Omega,\;\left|\omega-\omega_{n}^{\dagger}\right|\geq\varepsilon\}\geq\delta_{\varepsilon}.
\]
\item The exact bootstrap coverage curve uniformly approximates the population
coverage curve at rate $a_{n}$:
\[
\sup\{\left|c_{n,\alpha}^{*}(\omega)-c_{n,\alpha}(\omega)\right|:\omega\in\Omega\}=O_{p}(a_{n}).
\]
\item On $N_{n}$, $c_{n,\alpha}(\omega)$ is continuously differentiable
and locally invertible, and there exists a constant $\eta>0$ such
that, for all sufficiently large $n$,
\[
\inf\{\left|\partial\omega c_{n,\alpha}(\omega)\right|:\omega\in N_{n}\}\geq\eta.
\]
\item With probability tending to one, the bootstrap equation $c_{n,\alpha}^{*}(\omega)=1-\alpha$
has at least one solution in $N_{n}$. The calibrated learning rate
$\widehat{\omega}_{n}$ is any measurable selection from these local
bootstrap roots. 
\end{enumerate}
\end{assumption}

\section{Validity of the Root of the Exact Bootstrap Coverage Equation}

The next result separates bootstrap coverage approximation from inversion.
Once uniform approximation holds, the exact bootstrap root inherits
its rate.

\begin{theorem}[Root stability under uniform bootstrap coverage approximation]

Suppose Assumption 2 holds. Let $\widehat{\omega}_{n}$ be any solution
of $c_{n,\alpha}^{*}(\omega)=1-\alpha$ that lies within $N_{n}$.
Then, $\widehat{\omega}_{n}-\omega_{n}^{\dagger}=O_{p}(a_{n})$. Moreover,
\[
c_{n,\alpha}(\widehat{\omega}_{n})-(1-\alpha)=O_{p}(a_{n}).
\]
Thus, the exact bootstrap learning rate estimate tracks the level-specific
population root, in the sense that $\widehat{\omega}_{n}-\omega_{n}^{\dagger}=o_{p}(1)$.

\end{theorem}

The proof is provided in Section S4 of the Supplementary Material.

The rate $a_{n}$ is the uniform bootstrap coverage approximation
rate in Assumption 2(3). In smooth fixed-dimensional estimation problems,
$a_{n}$ may be $O(n^{-1})$ when a scalar containment statistic,
a deterministic equivalent posterior boundary, and the induced coverage
map admit uniformly second-order accurate bootstrap and posterior
quantile expansions. Otherwise, Theorem 1 remains valid, with $a_{n}$
denoting the rate delivered by the available uniform bootstrap approximation.

In smooth fixed-dimensional estimation problems, Assumption 2(3) can
be verified from the uniform bootstrap Edgeworth and posterior quantile
expansions.

To connect posterior Edgeworth corrections with frequentist coverage,
we restrict our attention to credible sets whose containment events
allow a scalar-boundary reduction. That is, for a fixed nominal level
$1-\alpha$, we assume that, uniformly for $\omega\in\Omega$, the
event $\{\theta_{0}\in C_{n,\omega,\alpha}\}$ is approximated, up
to an $o(n^{-1})$ error, by an event of the form $\{T_{n,\omega}\le r_{n,\alpha}(\omega)\}$.
Here, $T_{n,\omega}$ is a scalar containment statistic and $r_{n,\alpha}(\omega)$
denotes the deterministic equivalent boundary. Signed-pivot credible
intervals satisfy this reduction directly; the radial and quadratic-form
versions require a corresponding scalar-boundary expansion as an additional
assumption. This reduction is an assumption that allows data-dependent
corrections to the centre, scale, shape, and posterior quantile of
the credible set to be absorbed into either $T_{n,\omega}$ or into
the deterministic equivalent boundary, $r_{n,\alpha}(\omega)$. It
excludes  arbitrary highest posterior density sets unless their random
geometry allows a scalar deterministic equivalent representation.
Without this reduction, the joint expansion of the sampling pivot
and a random credible set boundary would be required.

The deterministic boundary expansion is where the posterior Edgeworth
expansion enters.

\begin{theorem}[Higher-order coverage expansion for scalar-boundary credible sets]

Fix $\alpha\in(0,1)$ and let $\Omega\subset(0,\infty)$ be compact.
Suppose that, uniformly for $\omega\in\Omega$, the containment event
of the credible set admits the scalar representation
\[
P_{0}(\{\theta_{0}\in C_{n,\omega,\alpha}\}\bigtriangleup\{T_{n,\omega}\le r_{n,\alpha}(\omega)\})=o(n^{-1}),
\]
where $T_{n,\omega}$ is a scalar containment statistic and $r_{n,\alpha}(\omega)$
denotes the deterministic equivalent boundary. The statistic $T_{n,\omega}$
may include corrections induced by the posterior centre, scale, and
shape. When the posterior centre, scale, or shape corrections are
absorbed into $T_{n,\omega}$, their contributions are captured by
$G_{1,\omega}$ and $G_{2,\omega}$, rather than only through the
boundary terms $\delta_{1,\alpha}$ and $\delta_{2,\alpha}$. If the
credible set boundary remains random, the theorem can be applied after
this reduction to the joint expansion of the pivot and random boundaries.
Let
\[
G_{n,\omega}(t)=P_{0}\{T_{n,\omega}\leq t\}.
\]
Assume that, uniformly for $\omega\in\Omega$ and for $t$ in an open
interval containing the relevant boundary values, 
\[
G_{n,\omega}(t)=G_{0,\omega}(t)+n^{-1/2}G_{1,\omega}(t)+n^{-1}G_{2,\omega}(t)+o(n^{-1}),
\]
where the remainder is uniform over this interval and $\omega\in\Omega$.
Suppose that $G_{0,\omega}$ has density $g_{0,\omega}$, $G_{1,\omega}$
has derivative $g_{1,\omega}$, and $g_{0,\omega}$ is continuously
differentiable. Assume that $G_{2,\omega}$ is uniformly equicontinuous
in the same neighbourhood. These derivatives are assumed to be uniformly
bounded and uniformly equicontinuous in the neighbourhood of the set
$\{r_{0,\alpha}(\omega):\omega\in\Omega\}$. In addition, $\{r_{0,\alpha}(\omega):\omega\in\Omega\}$
is contained in the interior of the interval in which the expansion
holds. Assume that the deterministic equivalent boundary satisfies
\[
r_{n,\alpha}(\omega)=r_{0,\alpha}(\omega)+n^{-1/2}\delta_{1,\alpha}(\omega)+n^{-1}\delta_{2,\alpha}(\omega)+o(n^{-1})
\]
for $\omega\in\Omega$, where $\sup_{\omega\in\Omega}(|\delta_{1,\alpha}(\omega)|+|\delta_{2,\alpha}(\omega)|)<\infty$.
Then, frequentist coverage expands as
\[
c_{n,\alpha}(\omega)=C_{0,\alpha}(\omega)+n^{-1/2}C_{1,\alpha}(\omega)+n^{-1}C_{2,\alpha}(\omega)+o(n^{-1}),
\]
for $\omega\in\Omega$, where $r_{0}=r_{0,\alpha}(\omega)$, $\delta_{1}=\delta_{1,\alpha}(\omega)$,
$\delta_{2}=\delta_{2,\alpha}(\omega)$, 
\[
C_{0,\alpha}(\omega)=G_{0,\omega}(r_{0}),\qquad C_{1,\alpha}(\omega)=G_{1,\omega}(r_{0})+g_{0,\omega}(r_{0})\delta_{1},
\]
and 
\[
C_{2,\alpha}(\omega)=G_{2,\omega}(r_{0})+g_{1,\omega}(r_{0})\delta_{1}+g_{0,\omega}(r_{0})\delta_{2}+\frac{1}{2}g'_{0,\omega}(r_{0})\delta_{1}^{2}.
\]

The term $G_{1,\omega}(r_{0})$ is the first correction for the distribution
of the chosen containment statistic; if $T_{n,\omega}$ is the uncorrected
sampling pivot, this is the sampling Edgeworth correction. The term
$g_{0,\omega}(r_{0})\delta_{1}$ is the first-order posterior correction
transmitted through the credible boundary. The second-order term contains
the second correction for the distribution of the selected containment
statistic, the interaction between the sampling density correction
and the first boundary correction, the second boundary correction,
and the quadratic effect of the first boundary correction. 

\end{theorem}

The proof is provided in Section S5 of the Supplementary Material.

\section{Implemented Stochastic Approximation}

The implemented GPC algorithm replaces $c_{n,\alpha}^{*}$ with a
Monte Carlo estimate and uses Robbins--Monro iteration. We write
the noisy estimation function as
\[
h_{n}(\omega_{t})+b_{t}+\xi_{t+1},\qquad h_{n}(\omega)=c_{n,\alpha}^{*}(\omega)-(1-\alpha),
\]
where $b_{t}$ is a predictable bias term and $\xi_{t+1}$ is the
martingale difference error with respect to the simulation filtration
$\mathcal{F}_{t}$. The bias term represents finite bootstrap, posterior
Monte Carlo, and boundary estimation errors. The following result
is local; global convergence requires additional monotonicity or recurrence
conditions.

\begin{corollary}[Algorithmic convergence]

Let
\[
h_{n}(\omega)=c_{n,\alpha}^{*}(\omega)-(1-\alpha).
\]
Suppose that, conditionally on the data and on a bootstrap-regular
event, $h_{n}$ is locally Lipschitz on the open interval $B_{n}\subset\Omega$,
that it has a unique zero $\widehat{\omega}_{n}\in B_{n}$, and that
$\widehat{\omega}_{n}$ is locally asymptotically stable for the ordinary
differential equation $\dot{\omega}=h_{n}(\omega)$. Let $K_{n}$
be a compact interval such that $\widehat{\omega}_{n}\in\textrm{int}(K_{n})$,
$K_{n}\subset B_{n}$, and $K_{n}$ is contained in the domain of
attraction for $\widehat{\omega}_{n}$. Assume $K_{n}=[l_{n},u_{n}]$
is chosen such that $h_{n}\left(l_{n}\right)>0$, $h_{n}(u_{n})<0$,
and $\widehat{\omega}_{n}$ is a unique equilibrium in $K_{n}$. The
localized recursion is defined as follows: 
\[
\omega_{t+1}^{K}=\Pi_{K_{n}}\left[\omega_{t}^{K}+\rho_{t}\{h_{n}(\omega_{t}^{K})+b_{t}+\xi_{t+1}\}\right],
\]
where $\Pi_{K_{n}}$ denotes projection onto $K_{n}$. Assume that
$\omega_{1}^{K}\in K_{n}$, that $b_{t}$ is predictable, and that
$\xi_{t+1}$ is a martingale difference error satisfying $E^{*}(\xi_{t+1}\mid\mathcal{F}_{t})=0$
and $\sup_{t}E^{*}(\xi_{t+1}^{2}\mid\mathcal{F}_{t})\le K_{n}^{'}$
for a finite $K_{n}^{'}$ on a regular event. Assume also that $\sum_{t=1}^{\infty}\rho_{t}=\infty$,
$\sum_{t=1}^{\infty}\rho_{t}^{2}<\infty$, and $\sum_{t=1}^{\infty}\rho_{t}|b_{t}|<\infty$.
Then $\omega_{t}^{K}\to\widehat{\omega}_{n}$ in $P^{*}$-probability,
conditional on the data, on the regular event. If the original projected
recursion on $\Omega$ coincides with the localized recursion and
remains in $K_{n}$ with conditional probability tending to one, the
original recursion converges to $\widehat{\omega}_{n}$ in $P^{*}$-probability.
With the plus-sign update, a simple sufficient condition for local
asymptotic stability is $(\omega-\widehat{\omega}_{n})h_{n}(\omega)<0$
for all $\omega$ in a neighbourhood of $\widehat{\omega}_{n}$, $\omega\ne\widehat{\omega}_{n}$.
In particular, if $h_{n}$ is differentiable at the root, then $\partial_{\omega}h_{n}(\widehat{\omega}_{n})<0$
is sufficient. If the bootstrap coverage curve is increasing at the
root, the sign of the stochastic approximation update must be reversed.

\end{corollary}

The proof is provided in Section S6 of the Supplementary Material.

\section{Discussion}

There are two aspects of the calibration problem. At a fixed nominal
level, bootstrap calibration is a valid method for estimating the
root under the same type of regularity required for the bootstrap
coverage approximation. However, across all nominal levels, the scalar
learning rate has a structural limitation. Unless the sampling covariance
and the posterior covariance are proportional, it cannot transform
the elliptical posterior into the correct sampling shape. This conclusion
is consistent with broader findings on covariance matching in \citet{Miller2021}.

Several extensions of this method would be useful. First, the stochastic
approximation results should be combined with a full analysis of the
errors due to finite bootstrap resampling, posterior Monte Carlo,
and boundary estimation. Second, a genuinely multivariate posterior
Edgeworth theorem for arbitrary highest posterior density sets would
remove the restrictions imposed by the scalar radii used here. Third,
non-smooth Gibbs losses require different expansions because first-order
normality may hold without the derivatives used in formal Edgeworth
series. Fourth, dependent data require a block or dependent multiplier
bootstrap argument. Finally, a high-dimensional calibration cannot
be obtained by using a fixed-dimensional Edgeworth expansion without
additional structure.

\section*{Declaration of the use of generative AI and AI-assisted technologies}

During the preparation of this work, the author used OpenAI ChatGPT
for English-language editing. After using this tool, the author reviewed
and edited the content as necessary and takes full responsibility
for the content of the publication.

\section*{Supplementary Material}

The detailed assumptions and proofs are provided in the accompanying
supplementary material.

\bibliographystyle{apalike2}
\bibliography{reference}



\section*{Supplementary material}

# Supplementary Material for "A Theory of Bootstrap Coverage Calibration for Generalized Posterior Credible Sets"

Masahiro Tanaka*

June 24, 2026

## S1 Detailed Assumptions

This supplement identifies a set of sufficient regularity conditions, supplemented by the stated uniform Edgeworth assumptions, under which the high-level statements in the main paper hold. These conditions are deliberately standard and can be replaced by problem-specific alternatives.

Let $R_n(\theta) = n^{-1}\sum_{i=1}^{n}\ell_\theta(X_i)$ and $R(\theta) = P_0\ell_\theta$. The population target $\theta_0$ is a unique interior minimizer of $R$. The criterion is six times continuously differentiable in a neighbourhood of $\theta_0$, its derivatives have an integrable envelope of sufficiently high order, and $J = \nabla^2 R(\theta_0)$ is positive definite. The prior density is positive and four times continuously differentiable near $\theta_0$. The estimator $\widehat{\theta}_n$ is a consistent interior local minimizer satisfying $\nabla R_n(\widehat{\theta}_n) = 0$. The posterior mass outside any fixed neighbourhood of $\widehat{\theta}_n$ is exponentially small, and the local Laplace expansion can be differentiated term by term. The same conditions hold for the bootstrap, conditional on the data, with $P^*$-probability tending to one. Finally, the relevant sampling pivots, bootstrap pivots, and posterior standardized contrasts possess uniform Edgeworth expansions on compact subsets containing the quantiles used to calibrate the posterior, to the same order in the density expansion and cumulants as in the regular Bayesian case. See Kolassa & Kuffner (2020) for further details.

## S2 Posterior Edgeworth and Conditional Local Expansions

For every compact interval $\Omega \subset (0,\infty)$ and every fixed scalar contrast $a^\top\theta$, the posterior distribution under $\Pi_{n,\omega}(\cdot \mid X^n)$ admits a uniform Edgeworth expansion for $\omega \in \Omega$. Writing

$$Z_{a,n} = \frac{a^\top\theta - m_{a,n}}{s_{a,n}},$$

where $m_{a,n}$ and $s_{a,n}$ are the posterior centre and standard deviation of $a^\top\theta$, the posterior density and distribution functions satisfy

$$f_{a,n}(z) = \phi(z)\left\{1 + \frac{\kappa_{3,a,n,\omega}}{6}H_3(z) + \frac{\kappa_{4,a,n,\omega}}{24}H_4(z) + \frac{\kappa^2_{3,a,n,\omega}}{72}H_6(z)\right\} + O_p(n^{-3/2}),$$

*Faculty of Economics, Fukuoka University, Fukuoka, Japan. Address: 8-19-1, Nanakuma, Jonan, Fukuoka, Japan 814-0180. E-mail: m.tanaka.tt@fukuoka-u.ac.jp.

and

$$F_{a,n}(z) = \Phi(z) - \phi(z)\left\{\frac{\kappa_{3,a,n,\omega}}{6}H_2(z) + \frac{\kappa_{4,a,n,\omega}}{24}H_3(z) + \frac{\kappa_{3,a,n,\omega}^2}{72}H_5(z)\right\} + O_p(n^{-3/2}),$$

uniformly for $z$ in compact sets and for $\omega \in \Omega$. Here, $H_j$ are probabilists' Hermite polynomials. The standardized cumulants satisfy

$$\kappa_{3,a,n,\omega} = O_p(n^{-1/2}), \qquad \kappa_{4,a,n,\omega} = O_p(n^{-1}),$$

as in the regular posterior cumulant theory of Kolassa & Kuffner (2020), adapted to the local expansion of a generalized posterior. The same expansion holds conditionally for the bootstrap posterior $\Pi^*_{n,\omega}(\cdot \mid X^{*n}, X^n)$, where $m_{a,n}, s_{a,n}, \kappa_{3,a,n,\omega}, \kappa_{4,a,n,\omega}$ are replaced by their bootstrap analogues $m^*_{a,n}, s^*_{a,n}, \kappa^*_{3,a,n,\omega}, \kappa^*_{4,a,n,\omega}$, with the remainder $O_{P^*}(n^{-3/2})$ uniformly for $\omega \in \Omega$ on bootstrap-regular events whose conditional probability tends to one. Consequently, posterior skewness perturbs smooth posterior quantiles and scalar credible boundaries at order $n^{-1/2}$, whereas posterior kurtosis and squared skewness contribute to order $n^{-1}$.

Let $\widehat{\theta}^*_n$ denote the corresponding interior local minimizer of $R^*_n$, with

$$\nabla R^*_n(\widehat{\theta}^*_n) = 0, \qquad \nabla^2 R^*_n(\widehat{\theta}^*_n) \to_{P^*} J,$$

where $J$ is positive definite, and

$$n^{1/2}(\widehat{\theta}_n - \theta_0) \rightsquigarrow N(0, \Psi).$$

For the bootstrap criterion, the local differentiability, curvature, and stochastic expansion conditions hold, conditional on the data, on events whose conditional probability tends to one. In addition, conditional on the data, $n^{1/2}(\widehat{\theta}^*_n - \widehat{\theta}_n)$ admits the bootstrap analogue of the sampling Edgeworth expansion for $n^{1/2}(\widehat{\theta}_n - \theta_0)$, uniformly for scalar pivots, as follows:

**Lemma 1** (Bootstrap local expansion)**.** *Under the preceding assumptions, conditional on the data and on an event whose $P^*$-probability tends to one, the bootstrap criterion admits the same form of local expansion as the original criterion, uniformly for $\theta$ in an $n^{-1/2}$-neighbourhood of the bootstrap minimizer $\widehat{\theta}^*_n$.*

*Proof.* Let $\theta = \widehat{\theta}^*_n + n^{-1/2}h$, where $h$ ranges over a fixed compact set. A Taylor expansion gives

$$\begin{aligned} n\{R^*_n(\widehat{\theta}^*_n + n^{-1/2}h) - R^*_n(\widehat{\theta}^*_n)\} &= n^{1/2}\nabla R^*_n(\widehat{\theta}^*_n)^\top h + \frac{1}{2}h^\top \nabla^2 R^*_n(\widehat{\theta}^*_n)h \\ &\quad + \frac{1}{6}n^{-1/2}\nabla^3 R^*_n(\widehat{\theta}^*_n)[h,h,h] \\ &\quad + \frac{1}{24}n^{-1}\nabla^4 R^*_n(\widehat{\theta}^*_n)[h,h,h,h] + r^*_n(h). \end{aligned}$$

The first term is zero, because $\widehat{\theta}^*_n$ is an interior stationary point of the bootstrap criterion. Moreover, the bootstrap local minimizer satisfies $\widehat{\theta}^*_n - \widehat{\theta}_n = O_{P^*}(n^{-1/2})$ on bootstrap-regular events, and the derivative processes are conditionally stochastically equicontinuous in a neighbourhood of $\theta_0$. Conditional uniform laws of large numbers for the derivatives imply that, for every fixed $M$,

$$\sup_{\|h\| \le M} \|\nabla^k R^*_n(\widehat{\theta}^*_n + n^{-1/2}h) - \nabla^k R_n(\widehat{\theta}_n + n^{-1/2}h)\| = o_{P^*}(1), \qquad k = 2,3,4,$$

after replacing the original derivative process with its bootstrap analogue and using the $O_{P^*}(n^{-1/2})$ separation between the two local centres. The sixth-derivative envelope controls the Taylor remainder uniformly. Thus, the bootstrap expansion has the same quadratic, cubic, and quartic forms as the original expansion, with the coefficients replaced by their bootstrap analogues. As the bootstrap Hessian converges conditionally to the same nonsingular limit, the expansion is valid on a bootstrap-regular event whose conditional probability tends to one. □

# S3 Proof of Proposition 1

*Proof.* Let

$$T_n = n(\widehat{\theta}_n - \theta_0)^\top J(\widehat{\theta}_n - \theta_0).$$

Using the continuous mapping theorem and the sampling limit in the main paper,

$$T_n \rightsquigarrow Q_\Lambda = \sum_{j=1}^{d} \lambda_j Z_j^2,$$

where $\lambda_1, ..., \lambda_d$ are the eigenvalues of $J^{1/2}\Psi J^{1/2}$. The posterior covariance expansion gives the ellipsoidal credible set

$$E_{n,\omega,\alpha} = \{\theta : n(\theta - \widehat{\theta}_n)^\top J(\theta - \widehat{\theta}_n) \le \frac{q_{d,1-\alpha}}{\omega} + o_p(1)\}.$$

Therefore

$$P_0(\theta_0 \in E_{n,\omega,\alpha}) = P_0\{T_n \le \frac{q_{d,1-\alpha}}{\omega} + o_p(1)\} \to P\{Q_\Lambda \le \frac{q_{d,1-\alpha}}{\omega}\},$$

provided that the limit distribution is continuous at the relevant point.

If a scalar $\omega_0$ calibrates every nominal level, then for all $x > 0$ of the form $x = q_{d,1-\alpha}$,

$$P\{Q_\Lambda \le \frac{x}{\omega_0}\} = P\{\chi_d^2 \le x\}.$$

As these quantiles range over $(0, \infty)$, $Q_\Lambda$ and $\omega_0^{-1}\chi_d^2$ follow the same distribution. Their Laplace transforms must agree:

$$\prod_{j=1}^{d}(1 + 2s\lambda_j)^{-1/2} = \left(1 + \frac{2s}{\omega_0}\right)^{-d/2}, \qquad s \ge 0.$$

The poles, or equivalently, the coefficients of the logarithmic expansion around zero, identify the multiset of eigenvalues; hence, $\lambda_1 = \cdots = \lambda_d = \omega_0^{-1}$. Conversely, if all the eigenvalues equal $\omega_0^{-1}$, then $Q_\Lambda \overset{d}{=} \omega_0^{-1}\chi_d^2$, and the nominal level is calibrated. □

# S4 Proof of Theorem 1

*Proof.* Let

$$\Delta_n = \sup_{\omega \in \Omega} |c^*_{n,\alpha}(\omega) - c_{n,\alpha}(\omega)|.$$

From Assumption 2(3), $\Delta_n = O_p(a_n)$. Work on the event, whose probability tends to one, on which a bootstrap root $\widehat{\omega}_n \in N_n$ exists and the uniform approximation bound holds.

The local separation condition implies consistency. Indeed, for any sufficiently small fixed $\varepsilon > 0$, Assumption 2(2) yields a $\delta_\varepsilon > 0$ such that, for all large $n$,

$$\inf_{\omega \in \Omega:\ |\omega - \omega_n^\dagger| \geq \varepsilon} |c_{n,\alpha}(\omega) - (1-\alpha)| \geq \delta_\varepsilon.$$

As $c^*_{n,\alpha}(\widehat{\omega}_n) = 1 - \alpha$,

$$|c_{n,\alpha}(\widehat{\omega}_n) - c^*_{n,\alpha}(\widehat{\omega}_n)| \leq \Delta_n.$$

Because $\Delta_n = o_p(1)$, the two preceding displays imply that

$$\widehat{\omega}_n - \omega_n^\dagger = o_p(1).$$

Next, we obtain the rate. Because

$$c_{n,\alpha}(\omega_n^\dagger) = 1 - \alpha, \qquad c^*_{n,\alpha}(\widehat{\omega}_n) = 1 - \alpha,$$

it follows that

$$\begin{aligned} |c_{n,\alpha}(\widehat{\omega}_n) - c_{n,\alpha}(\omega_n^\dagger)| &= |c_{n,\alpha}(\widehat{\omega}_n) - c^*_{n,\alpha}(\widehat{\omega}_n)| \\ &\leq \Delta_n. \end{aligned}$$

Because $N_n$ is convex, and $\widehat{\omega}_n$ and $\omega_n^\dagger$ lie in $N_n$, the mean value theorem yields

$$c_{n,\alpha}(\widehat{\omega}_n) - c_{n,\alpha}(\omega_n^\dagger) = \partial_\omega c_{n,\alpha}(\tilde{\omega}_n)(\widehat{\omega}_n - \omega_n^\dagger)$$

for some $\tilde{\omega}_n \in N_n$. By Assumption 2(4),

$$|\partial_\omega c_{n,\alpha}(\tilde{\omega}_n)| \geq \eta.$$

Therefore,

$$\eta|\widehat{\omega}_n - \omega_n^\dagger| \leq |c_{n,\alpha}(\widehat{\omega}_n) - c_{n,\alpha}(\omega_n^\dagger)| \leq \Delta_n,$$

and hence

$$|\widehat{\omega}_n - \omega_n^\dagger| \leq \frac{\Delta_n}{\eta} = O_p(a_n).$$

Finally, the coverage error follows directly from the identity for the bootstrap root:

$$\begin{aligned} |c_{n,\alpha}(\widehat{\omega}_n) - (1-\alpha)| &= |c_{n,\alpha}(\widehat{\omega}_n) - c^*_{n,\alpha}(\widehat{\omega}_n)| \\ &\leq \sup_{\omega \in \Omega} |c^*_{n,\alpha}(\omega) - c_{n,\alpha}(\omega)| \\ &= O_p(a_n). \end{aligned}$$

This proves both assertions. □

# S5 Proof of Theorem 2

*Proof.* Using a scalar representation,

$$c_{n,\alpha}(\omega) = G_{n,\omega}\{r_{n,\alpha}(\omega)\} + o(n^{-1})$$

for $\omega \in \Omega$. Write

$$r_{n,\alpha}(\omega) = r_0 + n^{-1/2}\delta_1 + n^{-1}\delta_2 + o(n^{-1}).$$

A Taylor expansion around $r_0$ yields

$$G_{0,\omega}\{r_{n,\alpha}(\omega)\} = G_{0,\omega}(r_0) + g_{0,\omega}(r_0)\{n^{-1/2}\delta_1 + n^{-1}\delta_2\} + \frac{1}{2}g'_{0,\omega}(r_0)n^{-1}\delta_1^2 + o(n^{-1}),$$

$$n^{-1/2}G_{1,\omega}\{r_{n,\alpha}(\omega)\} = n^{-1/2}G_{1,\omega}(r_0) + n^{-1}g_{1,\omega}(r_0)\delta_1 + o(n^{-1}),$$

and

$$n^{-1}G_{2,\omega}\{r_{n,\alpha}(\omega)\} = n^{-1}G_{2,\omega}(r_0) + o(n^{-1}).$$

Adding these three expansions yields the stated result. Uniformity follows from the uniform Edgeworth expansion, the uniform boundary expansion, and the smoothness assumptions of the derivatives over the compact set of relevant boundary values. □

# S6 Proof of Corollary 1

*Proof.* Condition on the data and on a bootstrap-regular event. For the localized recursion,

$$\omega_{t+1}^K = \Pi_{K_n}\left[\omega_t^K + \rho_t\{h_n(\omega_t^K) + b_t + \xi_{t+1}\}\right].$$

The martingale noise satisfies the Kushner–Clark negligibility condition (Kushner & Clark, 1978). Indeed, because

$$\sum_{t=1}^{\infty} \rho_t^2 E^*(\xi_{t+1}^2 \mid \mathcal{F}_t) \le K'_n \sum_{t=1}^{\infty} \rho_t^2 < \infty,$$

the martingale series $\sum_{t=1}^{\infty} \rho_t \xi_{t+1}$ converges in $P^*$-probability, conditionally on the data. Hence, for every fixed $T > 0$,

$$\sup_{m\ge t:\ \sum_{j=t}^{m}\rho_j\le T} \left|\sum_{j=t}^{m} \rho_j \xi_{j+1}\right| \to 0$$

in $P^*$-probability as $t \to \infty$. The predictable bias is negligible on the same stochastic approximation time scale because $\sum_{t=1}^{\infty} \rho_t |b_t| < \infty$ implies that

$$\sup_{m\ge t:\ \sum_{j=t}^{m}\rho_j\le T} \left|\sum_{j=t}^{m} \rho_j b_j\right| \to 0.$$

Thus, the linearly interpolated localized recursion is an asymptotic pseudo-trajectory of the projected ordinary differential equation (ODE) associated with $\dot{\omega} = h_n(\omega)$ on $K_n$.

As $h_n$ is locally Lipschitz on $B_n$, the ODE is locally well-posed. By assumption, $K_n \subset B_n$ is contained in the domain of attraction of the locally asymptotically stable equilibrium $\widehat{\omega}_n$. Hence, every solution of the limiting ODE starting in $K_n$ converges to $\widehat{\omega}_n$, and the projection is inactive for dynamics sufficiently close to the equilibrium. The standard Kushner–Clark ODE theorem for stochastic approximation (Kushner & Clark, 1978), together with the Robbins–Siegmund convergence argument applied to a local Lyapunov function (Robbins & Siegmund, 1971), yields $\omega_t^K \to \widehat{\omega}_n$ in $P^*$-probability, conditional on the data.

Finally, suppose that the original projected recursion on $\Omega$ remains in $K_n$ with conditional probability tending to one. On this event, the projection onto $K_n$ is inactive, and the original recursion coincides with the localized recursion. Therefore, the original recursion converges to $\widehat{\omega}_n$ in $P^*$-probability. This proves the corollary. □



\end{document}